\documentclass[%
 reprint,
 amsmath,amssymb,
 aps,
]{revtex4-2}

\usepackage{graphicx}% Include figure files
\usepackage[usenames]{color}
\usepackage{dcolumn}% Align table columns on decimal point
\usepackage{bm}% bold math
\begin{document}

%\preprint{APS/123-QED}

\title{Measuring cosmic expansion with diffractive gravitational scintillation of nanoHertz gravitational waves}% Force line breaks with \\

\author{Dylan L. Jow}%
 \email{djow@physics.utoronto.ca}
\affiliation{Canadian Institute for Theoretical Astrophysics, University of Toronto, 60 St. George Street, Toronto, ON M5S 3H8, Canada}
\affiliation{Department of Physics, University of Toronto, 60 St. George Street, Toronto, ON M5S 1A7, Canada}
\affiliation{Dunlap Institute for Astronomy \& Astrophysics, University of Toronto, AB 120-50 St. George Street, Toronto, ON M5S 3H4, Canada}

\author{Ue-Li Pen}
\affiliation{
Institute of Astronomy and Astrophysics, Academia Sinica, Astronomy-Mathematics Building, No. 1, Section 4,
Roosevelt Road, Taipei 10617, Taiwan
}
\affiliation{Canadian Institute for Theoretical Astrophysics, University of Toronto, 60 St. George Street, Toronto, ON M5S 3H8, Canada}
\affiliation{Department of Physics, University of Toronto, 60 St. George Street, Toronto, ON M5S 1A7, Canada}
\affiliation{Dunlap Institute for Astronomy \& Astrophysics, University of Toronto, AB 120-50 St. George Street, Toronto, ON M5S 3H4, Canada}
\affiliation{Perimeter Institute for Theoretical Physics, 31 Caroline St. North, Waterloo, ON, Canada N2L 2Y5}
\affiliation{Canadian Institute for Advanced Research, CIFAR program in Gravitation and Cosmology}

\date{\today}% It is always \today, today,

\begin{abstract}
The recent discovery of ultra-long wavelength gravitational waves through the advent of pulsar timing arrays (PTA) has opened up new avenues for fundamental science. Here we show that every PTA source will be diffractively lensed by potentially hundreds of galactic disks transverse to its line of sight, leading to modest modulations in the strain, $\Delta h / h \sim 10^{-3} \lambda^{-1}_{1 \rm pc.}$, due to wave lensing effects. The induced interference, or scintillation, pattern will be resolvable by coherent PTAs and may be leveraged, alongside fore-ground redshift information, to make precise measurements of cosmic expansion. If future PTA experiments can achieve enough signal-to-noise to detect these small modulations, hundreds of redshift-distance pairs may be inferred from the diffractive lensing of an individual PTA source. 
\end{abstract}

% ADD KEYWORDS: FRB, SCINTILLOMETRY, HUBBLE CONSTANT

\maketitle

%\tableofcontents

\section{Introduction}

The discovery of gravitational waves in ground-based interferometers and, most recently, in pulsar timing arrays (PTAs) \citep{Nanograv2023} will enable us to probe the sky on a wide range of wavelengths spanning many orders of magnitude. Gravitational waves are expected to exhibit wave lensing effects, providing new ways of probing dark matter structure on small-scales with millihertz gravitational waves \citep{Review, Savastano2023, TambaloZumalacarregui2023}. In this letter, we argue nanohertz gravitational wave sources will be diffractively lensed by hundreds of edge-on galaxies towards any given line of sight. While the effect will be small (a sub-percent modulation in strain), it will be ubiquitous and the resulting interference, or scintillation, pattern will be resolvable by galactic DSA-2000 or SKA-era PTAs \cite{2009A&A...493.1161S, 2019BAAS...51g.255H}. If detectable, one may effectively measure hundreds of redshift-distance pairs for a single PTA source, leading to measurements of the Hubble constant to one-part-in-ten-thousand. Notably, because of the large bending angles achieved by the diffractive gravitational lensing, the observed lensing time delays will be dominated by the geometric part, as opposed to the Shapiro delay. Thus, the proposed measurement does not rely on precise modeling of galaxy mass profiles, in contrast with previous strong lensing measurements \cite{2018SSRv..214...91S}. While a detection of the cumulative diffractive effect is futuristic, it may enable us to directly probe changes in the rate of cosmic expansion, as well as anisotropies in cosmic expansion.

\section{\label{sec:diffraction}Diffractive Gravitational Lensing}

Here we will briefly describe the diffractive lensing theory for a simple Gaussian lenses. First, let us consider the case of an isolated Gaussian lens with lens potential:
\begin{equation}
    \hat{\Psi}(\boldsymbol{\xi}) = A e^{-\frac{\xi^2_1}{2 \ell^2_1}} e^{-\frac{\xi^2_2}{2 \ell^2_2}},
\end{equation}
where $\boldsymbol{\xi} = [\xi_1, \xi_2]$ is the physical coordinate in the lens plane, and $\ell_1 \leq \ell_2$ set the angular size of the lens along the semi-minor and -major axes. Here we define the lens potential to be the Shapiro delay of the lens, so that the amplitude, $A$, has units of time, and for a lens of mass $M$ is of order $A \sim G M / c^3$. 

We want to compute the Kirchhoff-Fresnel integral which determines the complex amplification factor for the observed radiation field due to the lens. It is often convenient to define dimensionless coordinates $\boldsymbol{x} = \boldsymbol{\xi} / \ell_l$ and $\boldsymbol{y} = \boldsymbol{\eta} / \ell_l$, where $\boldsymbol{\eta}$ is the relative angular position between the source and lens, and we choose to normalize the angular coordinates by the smaller of the two scales, $\ell_1$, $\ell_2$. We also define the dimensionless lens potential $\Psi(\boldsymbol{x}) = \ell_1^{-2} D^{-1} \hat{\Psi}(\boldsymbol{x} \ell_1)$, where $D = D_l D_{ls} / D_s$ is a combination of the angular diameter distance to the lens, to the source, and between the lens and source. The Kirchhoff-Fresnel integral is then given by
\begin{equation}
    F(\boldsymbol{y}) = \frac{\nu}{2\pi i} \int \exp \left\{ i \nu \left[ \frac{1}{2} |\boldsymbol{x} - \boldsymbol{y}|^2 - \Psi(\boldsymbol{x}) \right]\right\} d^2{\boldsymbol{x}}
    \label{eq:KFI}
\end{equation}
where 
\begin{equation}
    \nu =(1 + z_l) \frac{\omega \ell^2 _1}{c \overline{d}},
\end{equation}
where $\omega$ is the angular frequency of the radiation incident at the observer and $z_l$ is the cosmological redshift of the lens. In the diffractive limit, $\omega \to 0$, we can evaluate the integral perturbatively:
\begin{equation}
    F({\boldsymbol{y}}) = 1 + i \nu \int \Psi(\boldsymbol{x}) e^{\frac{i \nu}{2} |\boldsymbol{x} - \boldsymbol{y}|^2} d^2\boldsymbol{x} + \mathcal{O}(\epsilon^2)
    \label{eq:generalperturbative}
\end{equation}
where the expansion parameter $\epsilon \equiv \kappa \nu$ is a product between the dimensionless frequency $\nu$ and the convergence, $\kappa \equiv |\frac{1}{2} \nabla^2_{\boldsymbol{\xi}} \hat{\Psi}(0)|$. The perturbative expansion is valid when $\epsilon \ll 1$ \cite{JowRegimes2023}. 

For simplicity, consider an off-axis source centred in the the semi-major axis of the lens, $\boldsymbol{y} = [y,0]$. The perturbative expansion is obtained analytically:
\begin{align}
    F(y) &= 1 + i \nu \kappa_r f(y, \nu) f(0, \nu s^2), \\
    f(y, \nu) &= \sqrt{\frac{\nu}{i + \nu}} \exp\{\frac{- \nu^2 y^2}{2 (1 + \nu^2)}\} \exp\{i \frac{\nu y^2}{2 (1 + \nu^2)}\},
    \label{eq:gaussianperturbatuve}
\end{align}
where we have defined $\kappa_1 = \frac{1}{2} \partial^2_{\xi_1} \hat{\Psi}(0)$ and $s = \ell_2 / \ell_1$. Note that the total convergence is bounded, $\kappa_1 \leq \kappa \leq \sqrt{2} \kappa_1$, so that as long as the convergence along the semi-minor axis satisfies $\kappa_1 \nu \ll 1$, then the diffractive limit holds. 

Let us note a few things about Eq.~\ref{eq:gaussianperturbatuve}. First, the diffractive modulation is exponentially suppressed whenever $y > \nu^{-1}$, setting an effective cross-sectional radius, $\xi_{\rm max} = \ell_1 \nu^{-1}$. Namely, in order for the lens to significantly modulate the flux from the source, the angular separation between the source and lens must be less than $\xi_{\rm max}$. This is equivalent to the familiar relation that a diffractive lens can bend light up to a maximum bending angle of $\alpha \sim \lambda / \ell_1$. Secondly, when the source is within this radius, the total flux from the lens -- defined a $\mu = |F - 1|^2$ -- is of order $\mu \sim \kappa^2_1 \nu^3 \sqrt{\nu^2 s^4 / (1 + \nu^2 s^4)}$. In the limit that $\nu s^2 \gg  1$, we obtain $\mu \sim \kappa^2_1 \nu^3$, which is simply the diffractive flux for a one-dimensional Gaussian lens. This condition is equivalent to $\ell_2 \gtrsim r_F$, where $r_F = \sqrt{\lambda \overline{d} / (1 + z_l)}$ is the Fresnel scale. Roughly speaking, the Fresnel scale sets the scale of the interference fringes that arise in wave optics. Thus, when one of the axes of the lens is large compared to the Fresnel scale, the lens behaves as a one-dimensional lens with convergence $\kappa_1$. The diffractive flux of a one-dimensional lens is always larger than the flux of an axisymmetric lens of equivalent width. 

Relating the dimensionless quantities to physical scales, we obtain
\begin{align}
    \nu &= 2 \pi \left( \frac{\ell_1}{r_F} \right)^2, \\
    \kappa_r &= \left( \frac{r_E}{\ell_1} \right)^2, \\
    \epsilon &\sim \nu \kappa_r = \frac{4 \pi R_s (1 + z_l)}{\lambda},
\end{align}
where $r_E = \sqrt{2 R_s \overline{d}}$ is the Einstein radius for a lens with Schwarzschild radius $R_s$. It immediately follows that the diffractive regime is attained when the wavelength of the radiation is large compared to the Schwarzschild radius of the lens. In this work, we will be interested in the cumulative diffractive flux of many Gaussian lenses. Since the diffractive regime is defined by a perturbative expansion, the fluxes simply add linearly. Thus, to compute the total diffractive flux, we simply add the flux of all the lenses for which the source falls within the lensing cross-section, $\sigma = 2 \pi \xi^2_{\rm max}$. In the next section we will perform this sum for nanohertz gravitational waves being diffracted by luminous galactic disks along the line of sight.

\section{\label{sec:galaxies}The Galactic Diffraction Grating}

Consider a gravitational wave source observed in a pulsar timing array (PTA) with a typical wavelength of $\lambda = 1\,{\rm pc}$. Galactic disks with typical masses on the order of $10^{10}\,M_\odot$ have Schwarzschild radii well below $1\,{\rm pc}$. Thus, galactic disks towards the direction of the gravitational wave will produce diffractive modulations. To obtain a quick estimate of the cumulative diffractive flux, we consider only the contribution from galaxies seen edge on. As noted in the previous section, highly asymmetric lenses produce larger diffractive fluxes. The Fresnel scale for a parsec-wavelength source at gigaparsec distances is $r_F \sim 10\,{\rm kpc}$, which coincides with the typical diameter of a galactic disk. Thus, edge-on galaxies are typically in the regime where $\ell_2 \gtrsim \theta_F$, so that they behave effectively as one-dimensional lenses. The individual diffractive flux from such a lens is given by
\begin{equation}
    \mu \sim \kappa^2_1 \nu^3 = 32 \pi^3 \left( \frac{R_s \ell_1}{\lambda r_F}  \right)^2. 
\end{equation}
An edge-on galaxy with $\ell_1 \sim 0.1\,{\rm kpc}$ yields $\mu \sim 10^{-7}$. This is, of course, a small number. However, in general, there will be a large number of edge-on galaxies contributing to the diffractive flux. In order for an edge-on galaxy with width $\ell_1$ to contribute, it needs to be within  $\xi_{\rm max}$ of the line-of-sight to the source. For an edge-on galaxy at a gigaparsec lensing a nanohertz gravitational wave, one obtains $\xi_{\rm max} \sim 1\,{\rm Mpc}$. In other words, edge-on galaxies within a megaparsec of the line-of-sight towards a PTA source will diffractively modulate the source. One arrives at the same conclusion by noting that a lens diffractively bends light up to an angle of $\alpha \sim \lambda / \ell_1 \sim 10'$.

Typically, there will be over $\sim 10^4$ galaxies within a megaparsec of any given line of sight out to a distance of $D_s \sim 1\,{\rm Gpc}$. If only one in a hundred of these galaxies is edge on (assuming galaxies have a typical aspect ratio of 1:100), hundreds of such galaxies will contribute significantly to the flux, meaning the cumulative diffractive flux is $\mu_{\rm tot.} \sim 10^{-5}$. However, as PTAs are sensitive directly to the strain of the gravitational radiation, the signal-to-noise depends on $\sqrt{\mu_{\rm tot.}} \sim 10^{-3} - 10^{-2}$. The size of this effect grows linearly with frequency so that at $f = 10^{-7}\,{\rm Hz}$, which is well within PTA sensitivities, the effect may grow above the percent level.

\begin{figure*}
    \centering
    \includegraphics[width=2\columnwidth]{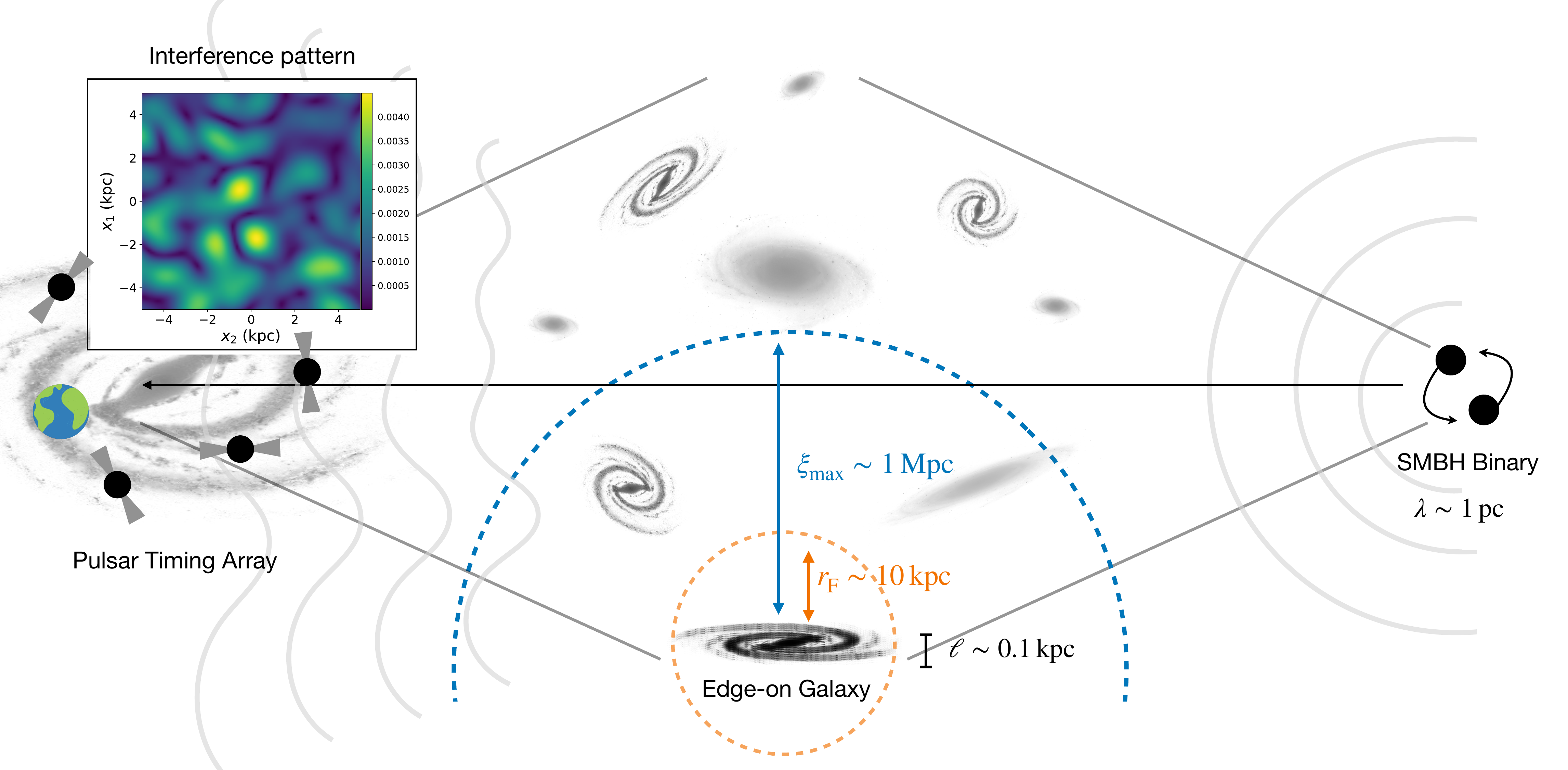}
    \caption{Diffractive lensing of a nanohertz gravitational wave by galactic disks. In geometric optics, the Einstein radius sets the optical depth of a lens, which for a typical galactic disk is only $\sim 10\,{\rm kpc}$. In the diffractive limit, the maximum bending angle is $\alpha \sim \lambda / \ell$, leading to a maximum impact parameter $\xi_{\rm max} \sim r^2_F / \ell \sim 1\,{\rm Mpc}$ for edge-on galaxies. Hundreds of edge-on galaxies may be within $1\,{\rm Mpc}$ of any given line of sight. The colour-map superimposed over the Milky Way shows the interference pattern imprinted onto the strain of the gravitational wave.}
    \label{fig:diffraction_grating_diagram}
\end{figure*}

\subsection{\label{sec:flux}Total integrated flux}

A more robust estimate of the cumulative diffractive effect can be obtained by integrating over the galaxy mass function, as well as possible orientations of the galaxies. Specifically, we will evaluate:
\begin{align}
\begin{split}
    &\mu_{\rm tot.}(z_s) = \int \frac{c dz_l}{(1+z_l) H(z_l)} dM d\iota  \\
    &\times 2 \pi \xi_{\rm max}^2(\lambda, \iota, z_l; z_s) \mu(M, \lambda, \iota, z_l; z_s) \phi(M, z_l),
    \label{eq:fluxintegral}
\end{split}
\end{align}
where $\phi(M, z_l)$ is the number density of galaxies with luminous mass $M$ at redshift $z_l$, and $\iota$ denotes the inclination angle of a galaxy relative to the radial direction from the line of sight to the source. The $c / (1 + z) H(z)$ term is $dR/dz_l$ where $dR$ is the proper distance. In principle, we should also integrate over the possible diameters and scale-heights of the galactic disks, but for simplicity we assume a fixed relationship between the radius of the galaxy and the disk mass, $R_{\rm gal} = a (M / 10^{10} M_\odot)^b $, where $a = 5.5\,{\rm kpc}$ and $b = 0.27$ \cite{Lange2016}. We will assume that every galaxy has a thin disk with a fixed aspect ratio, where the scale-height is one one-hundredth of the radius, $h_{\rm gal} = 0.1 R_{\rm gal}$, consistent with the observed value for the Milky Way \cite{2008ApJ...673..864J}.

Taking $\mu = \kappa^2_r \nu^3 \sqrt{\nu^2 s^4 / (1 + \nu^2 s^4)}$, we obtain 
\begin{align}
\begin{split}
    2 \pi \xi_{\rm max}^2 \mu 
    =  16 \pi^2 \frac{r_F^2 R_s^2}{\lambda^2} \sqrt{\frac{2 \pi \ell_r^2}{r_F^2 + 2 \pi \ell_r^2}},
\label{eq:fluxintegrand}
\end{split}
\end{align}
where $\ell_r$ is the projected width of the galaxy on the plane of the sky in the radial direction from the line of sight: $\ell_r = \max \left\{h_{\rm gal},  2 R_{\rm gal} \cos \iota \right\}$. For the galaxy stellar mass function, we adopt a double Schechter function with empirical parameters given by \citet{McLeod2021}. Now, performing the integral we obtain $\mu(z_s = 2) = 6 \times 10^{-6}$ for $\lambda = 1\,{\rm pc}$. The diffractive modulation of the strain is $\sqrt{\mu(z_s =2)} = 2 \times 10^{-3}$. The top panel of figure~\ref{fig:PTAdiffractive_H0SNR} shows the strain modulation as a function of wavelength. 

\section{\label{sec:H0}Measuring $H_0$ from diffractive lensing}

So far we have given a simple order-of-magnitude estimation of the collective diffractive effect of the galactic disks towards any given sight-line for a PTA source. We found that hundreds of edge-on galaxies may lead to just-sub-percent level total modulations of the observed strain. Such a small signal is not observable with present PTA experiments; a gravitational wave background was only very recently reported to $4 \sigma$ \cite{Nanograv2023}. However, let us consider what we might be able to learn by detecting this diffractive effect if and when it becomes possible in the future. We propose here that it is possible to measure cosmological parameters (specifically the Hubble constant) to a level of precision that would be difficult to achieve through any other means. The basic principle behind this idea is that if the cumulative lensing effect of hundreds of galaxies can be detected in a pulsar timing array, then, coupled with a wide-field redshift survey, one can, in principle, infer hundreds of redshift-distance pairs from a \textit{single} detection of a PTA source. In other words, one would be able to make extremely precise measurements of cosmic expansion from a single source. With multiple sources, one could potentially probe anisotropies in cosmic expansion to high precision. 

Cosmological measurements from lensing require precise measurements of the difference in time of arrival between the lensed images. \citet{JowRegimes2023} show that the diffractive modulation resulting from a single lens can be regarded as a result from the interference between the primary image and a weak diffractive image with a well-defined time delay. It is critical to note that in the diffractive regime the time-delay associated with the diffractive image is dominated by the geometric part of the delay. A lens can form a diffractive image out to $y \sim \nu^{-1}$. The total phase delay associated with the image is of order $\frac{\nu}{2} y^2 - \kappa \sim \nu^{-1} - \kappa$. Thus, in the diffractive limit ($\nu \kappa \ll 1$) the geometric part of the delay dominates.

Now, one can access this time delay directly by Fourier transforming the chromatic interference pattern as a function of frequency.  However, PTA sources are effectively monochromatic, and, therefore, in order to extract time-delay information one must rely on the spatial variation of the flux. To sketch how such a measurement might proceed, we must think of the pulsar timing array as an interferometer, as, indeed, it is. For simplicity, we will consider a weakly modulated source composed of many dim diffractive images, each with individual flux, $\mu_j \sim 10^{-7}$, and time delays, $\tau_j$. The response of the interferometer is given by
\begin{align}
    V &= 1 + \sum_j \mu^{1/2}_j e^{i \omega \tau_j},
\end{align}
where we have normalized the intensity of the images so that the primary image has unit flux. If the diffractive lenses are located at some angular position $\theta_j$ relative to the source's line-of-sight, we can further re-write the time delay for each diffractive image as the geometric time delay:
\begin{equation}
    \tau_j =  \frac{D_j \theta^2_j}{2c}.
    \label{eq:diffractivegeomtimedelay}
\end{equation}
For simplicity, we will assume that each image forms a delta function in delay space. Typically, however, wave effects smear the delta function out so that the flux from each lens arrives with a broader distribution of time delays. Nevertheless, the peak of this distribution is determined by the geometric delay (Eq.~\ref{eq:diffractivegeomtimedelay}).

Now let us assume that we have knowledge of the foreground towards the gravitational wave source we are observing. In particular, say we know the redshift of the source, $z_s$, each of the lensing galaxies, $z_j$, and also the angular position of the lensing galaxies, $\theta_j$, via some wide-field spectroscopic survey towards that line of sight. We can construct an estimate of the lensing time delays:
\begin{equation}
    \hat{\tau}_j (\hat{\Omega}) = \frac{D(z_j, z_s; \hat{\Omega}) \theta^2_j}{2c},
    \label{eq:tauhat}
\end{equation}
where $\hat{\Omega} = \{ \hat{H}_0, \hat{\Omega}_m, \hat{\Omega}_\Lambda \}$ are the cosmological parameters. The hats denote that these quantities represent variable guesses for the true values of the cosmological parameters, which we represent as $\Omega = \{ H_0, \Omega_m, \Omega_\Lambda\}$. The estimated time delay is equal to the actual time delay, $\hat{\tau}_j(\hat{\Omega}) = \tau_j$, when the chosen values of the cosmological parameters equal the true values, $\hat{\Omega} = \Omega$. The dependence of the time delay on $\hat{\Omega}$ comes through the fact that for a given set of cosmological parameters one can infer all of the lensing distances from the two redshifts, $z_j$, $z_s$. For low redshifts, the distance will most strongly depend on the Hubble constant, and so for simplicity we will consider the estimated time delay as a function of $\hat{H}_0$ alone, fixing $\hat{\Omega}_m = \Omega_m$, $\hat{\Omega}_\Lambda = \Omega_\Lambda$. Our goal is to measure the time delays of the diffractive images from the PTA and compare this to our estimated values to infer the cosmological parameters. However, because each of the individual diffractive images will generally be extremely dim it will be practically impossible to resolve the individual lensed images on the sky.

Instead, we propose a stacking analysis, treating the PTA as a phased array. By adding relative phase offsets to the different antennae, one can effectively point the telescope, selecting for the flux arriving at some particular angular location on the sky. Since we know, \textit{a priori}, where all the lenses are, we can simply ``point" our PTA at each lens, and measure the response. This requires that the angular resolution of the telescope is sufficient to resolve the individual lenses. Typical angular separations of the images will be $\sim 1'$.  For a galactic PTA with a baseline $\eta \sim 10\,{\rm kpc}$,the achievable angular resolution is set by $\theta \sim \lambda / \eta \sim 10''$ for $\lambda = 1\,{\rm pc}$. The angular resolution decreases with wavelength, and so the proposed measurement will fail for $\lambda > 10\,{\rm pc}$. Note that this requirement is equivalent to stating that the spatial variations in the interference pattern are small relative to the size of the PTA. Figure~\ref{fig:diffraction_grating_diagram} shows a simulation of the expected diffractive modulation for $\lambda = 1\,{\rm pc}$, where the interference fringes are typically a kiloparsec in size. 

Now, because the diffractive flux from each image is so small, the signal from each pointing will be indistinguishable from noise. Thus, we must stack the responses from each pointing to recover a signal, multiplying each component by a phase, $e^{-i \omega \hat{\tau}_j}$, determined by our estimate for the time delay for that lens. The idea is that if our guess for $H_0$ is correct, then our estimated delays will match the actual delays and, therefore, all of the components will be added in-phase with each other, leading to a large cumulative signal. If our estimated delays are incorrect, there will be no signal. The total response that results from this stacking procedure is given by
\begin{align}
    V_{\rm stack} = 1 + \sum_j \mu^{1/2}_j e^{i \hat{\phi}_j(\hat{H_0})},
    \label{eq:stackedresponse}
\end{align}
where each image is being added with a phase given by
\begin{equation}
    \hat{\phi}_j = \omega \left( \tau_j - \hat{\tau}_j(\hat{H}_0) \right).
\end{equation}
When $\hat{H}_0 = H_0$, then every $\hat{\phi}_j = 0$ and we get that $V_{\rm stack} = 1 + \sum_j \mu^{1/2}_j$. The point is that if we know the phase of the incoming images, then we can leverage a coherent PTA to directly sum the strain of the images. In principle, the maximum signal one can measure is the sum of strains, $\sum_j \mu^{1/2}_j$, which, in general, is much larger than the diffractive modulation we estimated in the previous section. In that section, we estimated the diffractive modulation as the square-root of the total flux, $\sqrt{\sum_j \mu_j}$. When there a hundred lenses, the coherent sum of strains may be an order of magnitude larger than $\sqrt{\sum_j \mu_j}$, so that the actual signal could potentially be much larger than what we predicted. However, in realistic diffractive lensing, the flux from a given lens does not arrive with a single well-defined time delay, but rather a range of delays centred around some value. As a result of this smearing, the amplitude of the stacked signal will be less than $\sum_j \mu^{1/2}_j$, and the square-root of the total flux is a better estimate of the signal strength. 

For a single lens, the response is periodic in $\hat{H}_0$ , as the response is also maximized whenever $\hat{\phi}_j = 2 n \pi$. Thus, a single lens is insufficient to infer the Hubble constant unambiguously. In general, multiple lenses will contribute and the response will be a sum of oscillatory functions. In fact, ${\rm Re} [V_{\rm stack} - 1]$ will be a sum of cosines:
\begin{equation}
     {\rm Re}\left[ V_{\rm stack}(\Delta H_0) - 1 \right] = \sum_j \mu^{1/2}_j \cos \left( 2 \pi \Pi^{-1}_j \Delta H_0 \right),
     \label{eq:stacked_resposne}
\end{equation}
where $\Pi_j = \lambda H_0 / c \tau_j$. This follows from the fact that at low red-shifts the angular diameter distances as scale as $D_l, D_s, D_{ls} \propto \hat{H}_0^{-1}$. It follows that the estimated delay (Eq.~\ref{eq:tauhat}) can be written as $\hat{\tau}_j = \tau_j (H_0 / \hat{H}_0)$, where $\tau_j$ and $H_0$ are the true values of the delay and the Hubble constant, respectively. Defining deviations from the fiducial value of the Hubble constant to be $\Delta H_0 = \hat{H_0} - H_0$, we find that for small deviations, the phase offset is given by
\begin{equation}
    \hat{\phi} = \omega (\tau - \hat{\tau}) \approx \frac{\omega \tau}{H_0} \Delta H_0. 
\end{equation}
Thus, for small deviations from $H_0$, the phase offset is linear in $\Delta H_0$, from which Eq.~\ref{eq:stacked_resposne} follows.

When the number of lenses is large enough, the response loses its periodic structure and one can unambiguously infer the Hubble constant from the value of $\hat{H}_0$ at which the signal peaks. The width of this peak, which sets the precision of our inference, is approximately
\begin{equation}
    \sigma_{H_0} = \frac{\lambda H_0}{c \langle \tau_j \rangle}.
    \label{eq:H0error}
\end{equation}
In general, this can be extremely small. For a lens with an impact parameter of $1\,{\rm Mpc}$ from the line of sight, halfway between the observer and the source at a distance of a gigaparsec, the time delay is roughly $\sim 10^3{\rm years}$. Thus, for a parsec-wavelength gravitational wave, one could potentially measure the Hubble constant to a precision of $\sigma_{H_0} / H_0 \sim 10^{-3}$; i.e. well below percent level.

% \begin{figure}
%     \centering
%     \includegraphics[width=0.5\columnwidth]{PTA_H0measure.pdf}
%     \caption[Stacked PTA response for many diffractive lenses]{The real part of the stacked response (equation~\ref{eq:stackedresponse}) for a $\lambda = 1\,{\rm pc}$ gravitational wave at a source redshift of $z_s = 2$, with $N_{\rm lens} = 50$ diffractive images formed by lenses uniformly distributed in a cylinder with proper distance, $dR = c dz / (1 + z) H(z)$, out to the source and radius $b = 1\,{\rm Mpc}$.}
%     \label{fig:PTAH0_mockmeasurement}
% \end{figure}

\begin{figure}
    \centering
    \includegraphics[width=\columnwidth]{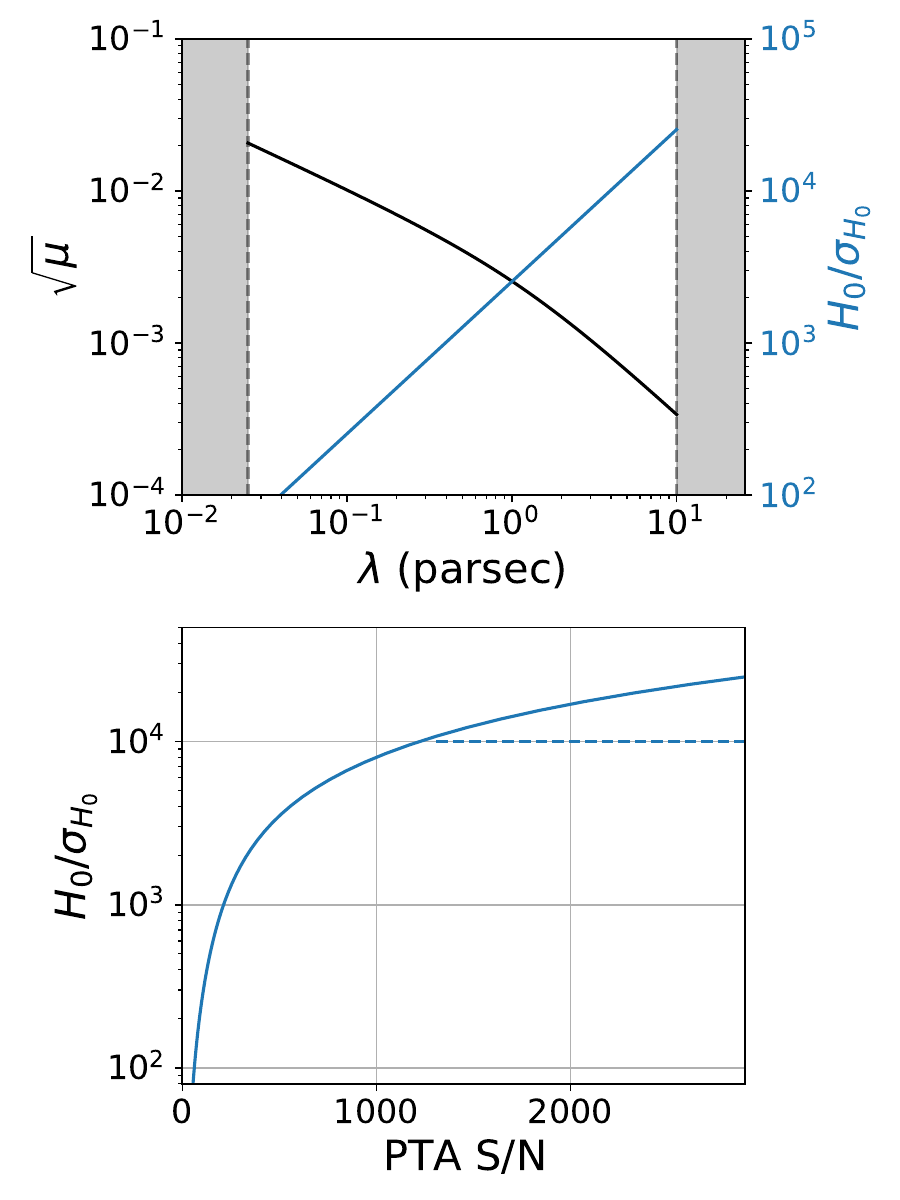}
    \caption[Precision on Hubble constant measurements from PTA diffractive modulations]{(Top) The black curve (negative slope) plotted against the left-hand y-axis is the cumulative diffractive modulation as a function of wavelength. The blue curve (positive slope) plotted against the right-hand y-axis is the achievable precision of Hubble constant, $H_0 / \sigma_{H_0}$. (Bottom) The two curves in the top panel plotted against each other, except the x-axis is given by the reciprocal of the total diffractive modulation, $1/\sqrt{\mu_{\rm tot.}}$, and is interpreted as the necessary signal-to-noise a PTA experiment would need to achieve to attain for a given precision on $H_0$. The dashed line shows the effective maximum precision that can be achieved due to the effect of lens' peculiar velocity.}
    \label{fig:PTAdiffractive_H0SNR}
\end{figure}

What is the optimal wavelength to attempt this measurement with? Pulsar timing arrays are sensitive to a large range of wavelengths. The precision on the Hubble constant increases linearly with wavelength; however, the cumulative flux decreases. While larger wavelengths mean larger lensing cross-sections, the actual flux from individual lenses decreases faster with wavelength. Figure~\ref{fig:PTAdiffractive_H0SNR} shows the estimated diffractive modulation (Eq.~\ref{eq:fluxintegral}) alongside the detection level for the Hubble constant, $H_0 / \sigma_{H_0}$ one would obtain at that wavelength. The grey region on the left represents an effective cut-off: for wavelengths below this, the number of galaxies that significantly contribute to the flux is too small to unambiguously determine $H_0$. The high wavelength cutoff is obtained when the PTA can no longer resolve the individual galaxies:  $\lambda / \eta = 1'$. The bottom panel of figure~\ref{fig:PTAdiffractive_H0SNR} shows the same information in a different way, plotting the $H_0$ precision curve in the right panel against the diffractive modulation curve. However, we have converted the diffractive modulation, $\sqrt{\mu_{\rm tot.}}$, to an effective signal-to-noise one would need to achieve with a PTA in order to detect such a small modulation. That is, if the modulation is $\sim 10^{-3}$ then one would need to be able to detect a single gravitational wave source with a signal-to-noise of more than $1000$ to detect such a modulation. However, if this signal-to-noise can be achieved, one could determine the Hubble constant to within a part in ten-thousand. This is an extreme level of precision that cannot be achieved with other probes.

At this level of precision, the effect of the lens galaxies' peculiar velocities play an important role, as the observed redshifts deviate from the cosmological redshift: $z^{\rm obs}_l = z_l + \delta z_l$. The contribution to the redshift from the peculiar velocity results in an error on the estimated phase of the lensed image: $\delta \phi_j = \delta z_l \phi_j / z_l$. When this error exceeds roughly $2\pi$-radians, the lens no longer contributes to the signal-to-noise of the Hubble constant. For typical peculiar velocities $\delta z_l / z_l \sim 10^{-3}$ \cite{2022MNRAS.514.4696W}, and since $H_0 / \sigma_{H_0} \sim \langle \phi_j \rangle$, the result is to place an effective upper-limit on the achievable precision of $H_0 / \sigma_{H_0} < 10^4$. 
 
Now assuming that the signal for the recent $4\sigma$ detection of a gravitational wave background is dominated by only a handful of bright super massive black hole mergers (following expectations of galaxy merger rates \citep{2014ApJ...789..156M, 2022ApJ...941..119B}), then at best the current achievable signal-to-noise is ${\rm S/N} \sim 10$. This is achieved with 67 pulsars. The signal-to-noise is expected to grow linearly with the number of pulsars. Given that there are tens-of-thousands of pulsars in the galaxy, the upper limit for the PTA signal-to-noise is larger than $S/N \sim 1000$. However, achieving precise timing models for that number of pulsars is extremely futuristic. More near term, we may expect next-generation radio telescopes such as the square-kilometre array (SKA) to add hundreds of pulsars to pulsar timing arrays \cite{2009A&A...493.1161S, 2015aska.confE..37J}. This would already be enough to perform precision cosmology measurements that are competitive with other probes. Achieving this signal-to-noise alone is insufficient for our proposed measurement. Galactic baselines between pulsar pairs are also needed, as well as precisely known distances to each pulsar in order to use the PTA as a coherent phased array. These conditions, while futuristic, are likely to be met by SKA or future SKA-like experiments.

\section{Conclusion}

Unlike most strong lensing effects where one must hope a given source happens to lie close enough to a lens  to detect anything, every ultra-long-wavelength gravitational wave source will exhibit diffractive modulation from the galactic diffraction grating. Moreover, since the time delay associated with diffractive images is dominated by the geometric delay, the effect is independent of the precise details of the potentials of the lensing galaxies. While the total effect will be small, by considering PTAs as phased arrays and utilizing foreground information to perform a stacking analysis, it may be detected. Such a detection can be used to make extremely precise measurements of the Hubble constant from a single source.

\bibliography{apssamp}% Produces the bibliography via BibTeX.

\end{document}